\documentclass[12pt]{article}
\usepackage{amssymb,amsmath} 
\usepackage{graphicx}
\usepackage{epsfig}
\usepackage{epstopdf}
\usepackage{latexsym}
\usepackage{psfrag}
\usepackage{subfigure}
\usepackage{booktabs}
\usepackage{braket}
\usepackage{bbm} 
\usepackage[toc]{appendix}
\usepackage{color,soul}
\usepackage{datetime}
\usepackage[
      colorlinks=true,
      linkcolor=darkblue,  
      urlcolor=blue,    
      filecolor=blue,     
      citecolor=pink,
      linktocpage=true,
      pdfstartview=FitV,
      bookmarksopen=true  
      ]{hyperref}

\definecolor{darkblue}{rgb}{0.2, 0, 0.8} 
\definecolor{pink}{rgb}{1, 0.15, 0.15}   

\numberwithin{equation}{section}

\newcommand{\req}[1]{(\ref{#1})} 
\newcommand{\labell}[1]{\label{#1}}

\newcommand{\bea}{\begin{eqnarray}}
\newcommand{\eea}{\end{eqnarray}}
\newcommand{\ba}{\begin{eqnarray}}
\newcommand{\ea}{\end{eqnarray}}

\newcommand{\beq}{\begin{equation}}
\newcommand{\eeq}{\end{equation} }
\newcommand{\beqa}{\begin{eqnarray}}
\newcommand{\eeqa}{\end{eqnarray}}
\newcommand{\beqar}{\begin{eqnarray*}}
\newcommand{\eeqar}{\end{eqnarray*}}

\renewcommand{\req}[1]{eq.~(\ref{#1})}
\newcommand{\twist}{\tau}
\newcommand{\reef}[1]{(\ref{#1})}
\newcommand{\ssc}{\scriptscriptstyle}
\newcommand{\eg}{{\it e.g.,}\ }
\newcommand{\ie}{{\it i.e.,}\ }

\newcommand{\cO}{\mathcal{O}}

\newcommand{\ren}{R\'enyi\ }

\newcommand{\see}{S_{\ssc EE}}

\newcommand{\ctt}{C_{\ssc T}}

\DeclareMathOperator{\tr}{Tr}

\renewcommand{\href}[2]{#2}

\begin{document}  


\begin{titlepage}

 \begin{flushright}
{\tt IFT-UAM/CSIC-15-083} \\
\end{flushright}

\vspace*{2.3cm}

\begin{center}
{\LARGE \bf Universal entanglement \\ for higher dimensional cones} \\

\vspace*{1.2cm}

{\bf Pablo Bueno$^{1}$ and Robert C. Myers$^{2}$}
\medskip

$^{1}$Instituto de F\'isica Te\'orica UAM/CSIC \\
C/ Nicol\'as Cabrera, 13-15, C.U. Cantoblanco, 28049 Madrid, Spain
\bigskip

$^{2}$Perimeter Institute for Theoretical Physics \\
31 Caroline Street North, ON N2L 2Y5, Canada 
\bigskip

p.bueno@csic.es, rmyers@perimeterinstitute.ca  \\
\end{center}

\vspace*{0.1cm}

\begin{abstract}  
The entanglement entropy of a generic $d$-dimensional conformal field theory receives a regulator independent contribution when the entangling surface contains a (hyper)conical singularity of opening angle $\Omega$, codified in a function $a^{\ssc (d)}(\Omega)$. In {\tt arXiv:1505.04804}, we proposed that for three-dimensional conformal field theories, the coefficient $\sigma^{\ssc (3)}$ characterizing the limit where the surface becomes smooth is proportional to the central charge $C_{\ssc T}$ appearing in the two-point function of the stress tensor. In this paper, we prove this relation for general three-dimensional holographic theories, and extend the result to general dimensions. In particular, we define a generalized coefficient $\sigma^{\ssc (d)}$ to characterize the almost smooth limit of a (hyper)conical singularity in entangling surfaces in higher dimensions. We show then that this coefficient is universally related to $C_{\ssc T}$ for general holographic theories and provide a general formula for the ratio $\sigma^{\ssc (d)}/C_{\ssc T}$ in arbitrary dimensions. We conjecture that the latter ratio is universal for general CFTs. Further, based on our recent results in {\tt arXiv:1507.06997}, we propose an extension of this relation to general R\'enyi entropies, which we show passes several consistency checks in $d=4$ and $6$.
\end{abstract}

\end{titlepage}

\setcounter{tocdepth}{2}
{\small
\setlength\parskip{-0.5mm} 
\tableofcontents
}

\section{Introduction}
\label{sec:Introduction}
\begin{figure}[h]
\center
 \subfigure[]{\label{pla}\includegraphics[scale=0.75]{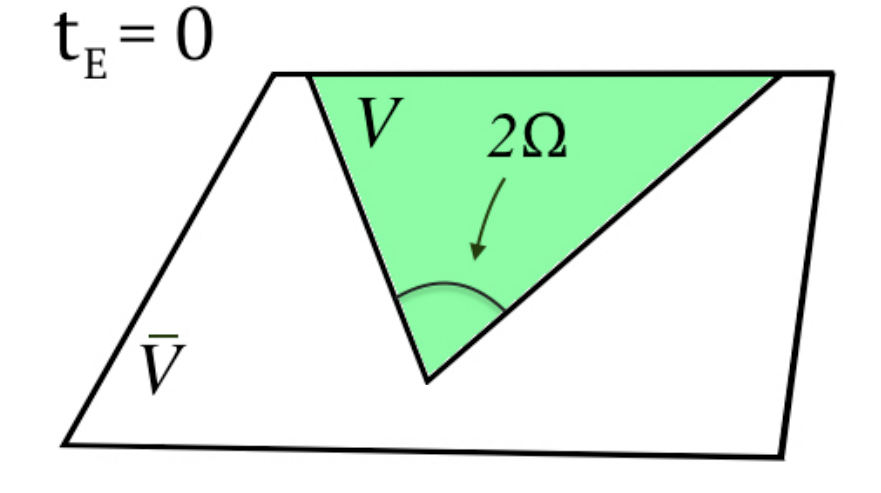}} \hspace{1.75cm} 
 \subfigure[]{\label{con} \includegraphics[scale=.43]{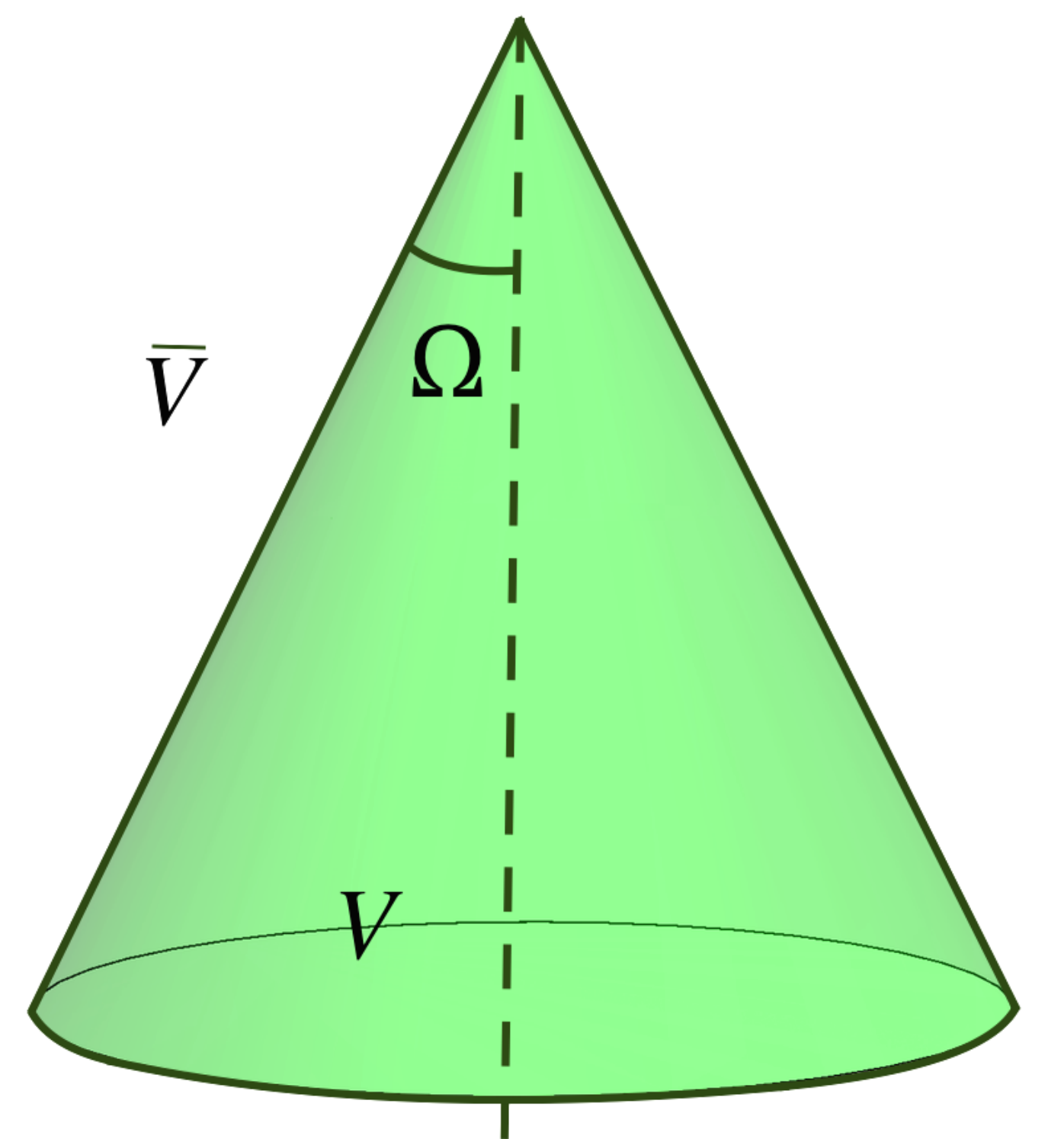}}  
\caption{In panel (a), we show an entangling region $V$ whose boundary contains a sharp corner of opening angle $2\Omega$. In panel (b), we show the analogous surface in $d=4$, \ie a region whose boundary contains a conical singularity of opening angle $\Omega$. The smooth limit is found in both cases for $\Omega\rightarrow \pi/2$.}    
\labell{cones} 
\centering
\end{figure}

Entanglement entropy (EE) and more generally \ren entropy has long been seen as an interesting probe of quantum field theories (QFTs), \eg \cite{cardy0,CHdir}. Typically in this context, one chooses some region $V$ on a Cauchy surface (\eg a constant time slice) and then evaluates the reduced density matrix $\rho_V$ by integrating out the degrees of freedom in the complementary region $\overline{V}$. The \ren and entanglement entropies are then defined as
\begin{align}\label{renyi}
S_n(V)=\frac{1}{1-n}\log\, \tr \rho_V^n \,,\qquad \see(V)=\lim_{n\to1}S_n(V)=-\tr \left( \rho_V \log \rho_V \right)\,.
\end{align} 
The calculation of these quantities must be regulated, \eg by a short distance cut-off $\delta$, because of an infinite number of short distance correlations in the vicinity of the entangling surface, \ie the boundary of $V$. The regulated results are dominated by various power law divergences, where the powers depend on the spacetime dimension $d$. While these divergent terms have an interesting geometric character \cite{grove,refine,sing} \eg the leading `area law' contribution: $S_n \simeq c_{d-2} \,{\cal A}(\partial V)/\delta^{d-2}$, the corresponding coefficients depend on the details of the regulator. However, examining $S_n$ and $\see$ in detail will also reveal universal contributions, whose coefficients are independent of the regulator and so provide unambiguous information about the underlying QFT. In particular, if the entangling surface is smooth, in an even number of dimensions, the universal contribution is characterized by a logarithmic divergence while in an odd number of dimensions, the constant contribution  (\ie $\delta$-independent term) will be universal if calculated with sufficient care \cite{mutual}. That is,
\begin{eqnarray}\labell{smooth}
S^{\rm univ}_n(V)=  \left\lbrace \begin{array}{cll}
\hspace{-1.7cm}(-1)^{\frac{d-1}{2}}\,s_n^{\rm univ}(V)\, && d\quad \text{odd}\, 
,\\  (-1)^{\frac{d-2}{2}}\, s_n^{\rm univ}(V) \, \log(R/\delta)&& d\quad \text{even}\, ,\end{array}\, \right.
\end{eqnarray}
where $R$ is some length scale characterizing the entangling region $V$.
However, the situation changes when the entangling surface $\partial V$ contains geometric singularities. In particular, with a conical singularity, as illustrated in Figure \ref{cones},  the regulator-independent terms take the form
\begin{eqnarray}\labell{cone}
S_n^{\rm univ}(V)= \left\lbrace \begin{array}{cll}
\hspace{-0.15cm}(-1)^{\frac{d-1}{2}}\, a_n^{\ssc (d)}(\Omega)\,\log(R/\delta)  && d \quad \text{odd}\, , \\ 
 (-1)^{\frac{d-2}{2}}\, a_n^{\ssc (d)}(\Omega)\,\log^2(R/\delta)&& d\quad \text{even}\, ,\end{array}\, \right.
\end{eqnarray}
where $a_n^{\ssc (d)}(\Omega)$ are functions of the opening angle $\Omega$ of the cone.\footnote{\label{foot}Please notice that as illustrated in Figure \ref{pla}, the opening angle of a corner in $d=3$ is defined to be `$2\Omega$' in the present paper. This contrasts with our conventions in \cite{bueno1,bueno3,bueno2}, where the same angle was called `$\theta$' and `$\Omega$,' respectively. The present convention simplifies the connection to higher-dimensional cones, for which we are using the same convention as in \cite{sing}. Notice, however, that the corner coefficient is defined in eq.~\reef{limi} so that $\sigma^{\ssc (3)}_n$ agrees with the coefficients evaluated in \cite{bueno1,bueno3,bueno2} for three dimensions.} The appearance of a new logarithmic divergence associated with sharp corners in the entangling surface is well-known for the case of three dimensions \cite{CHdir,hirata,fradkin}. The appearance of new universal terms in higher dimensions was first noted \cite{sing,igor7} using holographic entanglement entropy \cite{rt0}. Based on the latter results, our signs are chosen in \req{cones} to ensure that $a_n^{(d)}(\Omega)\geq 0$ for $0\le\Omega\le\pi$ for all $d$.

Now if the entanglement or \ren entropies are evaluated in a pure state, the results must be identical for the region $V$ and its complement $\overline{V}$. Therefore the corner function must satisfy
\beq
a_n^{\ssc (d)}(\Omega) = a_n^{\ssc (d)}(\pi-\Omega)\,.
\labell{pure}
\eeq
Further, our convention is that the entangling surface becomes smooth with $\Omega=\pi/2$ and hence $a_n^{\ssc (d)}(\Omega=\pi/2)=0$. Now assuming that these functions are smooth in the vicinity of $\Omega=\pi/2$, these two results constrain the form of the cone functions with
\begin{align}\labell{limi} 
a^{\ssc (d)}_n\left(\Omega\rightarrow \pi/2\right)=4\ \sigma^{\ssc (d)}_n\ \left(\frac{\pi}2-\Omega \right)^2\, ,
\end{align}
in general. Hence the universal corner contribution \reef{cone} defines a set of coefficients $\sigma_n^{\ssc (d)}$ which encode regulator-independent information about the underlying QFT. 

While the above comments apply for general QFTs, we will focus on conformal field theories (CFTs) throughout the following.
In refs.~\cite{bueno1,bueno2,bueno3}, we considered the properties of the corner coefficients $\sigma^{\ssc (3)}_n$ arising in three-dimensional CFTs. In particular, for the coefficient appearing in the EE, we argued that 
\begin{align}\labell{conj1}
\sigma^{\ssc (3)}\equiv\sigma^{\ssc (3)}_1 =\frac{\pi^2}{24}\,\ctt\, ,
\end{align}
for general three-dimensional CFTs. That is, $\sigma^{\ssc (3)}$ is proportional to the central charge $\ctt$ appearing in the two-point correlator of the stress tensor. Evidence for this conjecture comes from free scalars and fermions, as well as certain holographic theories \cite{bueno1,bueno2,Elvang}. We extend the proof to general holographic theories in the present paper --- see also \cite{rxm} for an alternate approach. Of course, it is tempting to think about possible extensions of the above relation to higher dimensions. In fact, an analogous result is known for general CFTs in $d=4$ where \cite{igor7}
\beq
\sigma^{\ssc (4)}=\frac{\pi^4}{640}\, \ctt  \,.
\eeq
In this paper, we prove that, at least for general holographic CFTs, $\sigma^{\ssc (d)}$ is indeed proportional to $\ctt$ in arbitrary dimensions. In particular, we find the general formula\footnote{Alternative proofs for general holographic theories in $d=3,\ 4$ and $6$ dimensions were presented in \cite{rxm}, using a different formalism. Some steps in this direction were also taken in \cite{alisha}.}
\begin{eqnarray}\label{CON}
\sigma^{\ssc (d)}=\ctt\ \frac{\pi^{d-1}(d-1)(d-2)\Gamma[\frac{d-1}{2}]^2}{8\,\Gamma[{d}/{2}]^2\,\Gamma[d+2]}\times\left\lbrace \begin{array}{cll}
\pi&& d \text{ odd}\, ,\\ 
1 && d\, \,\text{even} \,.\end{array}\, \right.
\end{eqnarray}
Hence, we conjecture that the cone coefficients $\sigma^{\ssc (d)}$ are related to $\ctt$ through \req{CON} for general CFTs. In Table \ref{tbl1}, we show the values $\sigma^{\ssc (d)}/\ctt$ for $d=3,4,\cdots,10$.

\begin{table*} 
  \centering
  \begin{tabular}{c||c|c|c|c|c|c|c|c} 
 $d$  &   3 & 4 & 5 & 6 &7&8&9&10\\
    \hline\hline \rule{0pt}{1.5em}
  $\sigma^{\ssc (d)}/\ctt$  & $\frac{\pi^2}{24}$ & $\frac{\pi^4}{640}$&   $\frac{\pi^4}{270}$ & $ \frac{\pi^6}{14336}$  & $\frac{\pi^6}{9450}$&$\frac{5\pi^8}{3538944}$&$\frac{4 \pi ^8}{2480625}$&$\frac{7\pi^{10}}{415236096}$\end{tabular}
  \caption{Cone coefficients $\sigma^{\ssc (d)}$ normalized by the stress tensor charge $\ctt$ in general holographic theories for various dimensions.}
\label{tbl1}   
\end{table*}
%
%


In \cite{bueno3}, we proposed a generalization of \req{conj1} to general R\'enyi entropies. According to this, the corresponding corner coefficients $\sigma_n^{\ssc (3)}$ are related to the scaling dimensions of the corresponding twist operators $h_n$ --- see Appendix \ref{Atwist} for definitions --- through
\begin{eqnarray}\labell{conj3a}
\sigma_n^{\ssc (3)}=\frac{1}{\pi}\,\frac{h_n}{n-1}\, .
\end{eqnarray}
We have verified that \req{conj3a} is satisfied for all integer values of $n$ and in the limit $n\rightarrow \infty$ both for a free scalar and a free fermion \cite{bueno3} --- see also \cite{Elvang,dowker}. Now all these results suggest a natural extension to the \ren cone coefficients in higher dimensions. In particular, the expansion  of the scaling dimension in general dimensions --- see eq.~\reef{expand} below --- suggests that  our result \reef{CON} extends to
 \begin{eqnarray}\labell{new9a}
\sigma_n^{\ssc(d)}=\frac{h_n}{n-1}\ \frac{(d-1)(d-2)\,\pi^{\frac{d-4}{2}}\,\Gamma\left[\frac{d-1}2\right]^2}{16\ \Gamma[{d}/{2}]^3}\ \times\,\left\lbrace 
\begin{array}{cll}
\pi && d\,\, \text{odd}\,,\\
1&& d\text{ even}\,.\end{array}\, \right.
\end{eqnarray}
We will show that this expression is consistent with previous results obtained for four- and six-dimensional CFTs \cite{aitor9,ben2}.

The remainder of the paper is organized as follows. In section \ref{coness}, we use the results of \cite{Mezei} to prove our generalized conjecture \reef{CON} for general holographic theories. In section \ref{renyi}, we extend this conjecture to the R\'enyi cone coefficients \reef{new9a} and use it to establish certain relations involving the structure of R\'enyi entropies for general entangling regions in four-dimensional theories. In section \ref{discuss}, we conclude with a brief discussion of our findings. Appendix \ref{Atwist} reviews some relevant information about twist operators and their conformal scaling dimensions.

\section{Cone coefficients for EE in general dimensions}
\label{coness}

Our approach to proving \req{CON} is to take advantage of the results in \cite{Mezei} with regards to the leading correction to the entanglement entropy for a slightly deformed sphere $S^{d-2}$. In \cite{Mezei,Mezei0}, the authors consider the case in which $\partial V$ is a slightly deformed $S^{d-2}$ sphere of radius $R$ parametrized in polar coordinates as
\beq \label{mezei1}
r(\Omega_{d-2})/R=1+\epsilon \sum_{\ell,m_1,...,m_{d-3}} a_{{\ell},m_1,...,m_{d-3}} Y_{\ell,m_1,...,m_{d-3}}(\Omega_{d-2})\, ,
\eeq
where $\epsilon$ is an infinitesimal parameter and $a_{\ell,m_1,\cdots,m_{d-3}}$ are some constant coefficients characterizing the deformation, Further,  $Y_{l,m_1,...,m_{d-3}}(\Omega_{d-2})$ are (real) hyper-spherical harmonics\footnote{Here as in \cite{Mezei}, the hyper-spherical harmonics are normalized such that:
\beq
\int d\Omega_{d-2}\ Y_{\ell,m_1,...,m_{d-3}}\,Y_{\ell^{\prime},m_1^{\prime},...,m_{d-3}^{\prime}}=\delta_{\ell\ell^{\prime}}\,\delta_{m_1m_1^{\prime}}\cdots\, \delta_{m_{d-3}m_{d-3}^{\prime}}\,.
\eeq} on $S^{d-2}$ with coordinates $\Omega_{d-2}$. In particular, they are eigenfunctions of the Laplacian on the sphere with
\beq \labell{hypers}
\triangle_{S^{d-2}}\, Y_{\ell,m_1,...,m_{d-3}}(\Omega_{d-2})=-\ell (\ell +d-3)\, Y_{\ell,m_1,...,m_{d-3}}(\Omega_{d-2})\, .
\eeq
For such a deformed sphere, the coefficient in the universal part of the EE \reef{smooth} takes the form
\beq\labell{defo}
s^{\rm univ}(V)=s^{\ssc (d)}_{\rm sphere}+\epsilon^2\, s^{\ssc (d)}_{2}(V)+\mathcal{O}(\epsilon^3)\, .
\eeq
Of course, the leading term here corresponds to the universal contribution for the undeformed sphere, which is given by \cite{ctheorem} 
\beq\labell{spa}
s^{\ssc (d)}_{\rm sphere}=\left\lbrace \begin{array}{cll}
F 
&& d\quad \text{odd}\, , \\ 4\,A && d \quad \text{even}\, .
\end{array}\, \right.\, 
\eeq
In even dimensions, $A$ is precisely the coefficient appearing in the A-type trace anomaly while in odd dimensions, $F$ can be identified with the universal contribution in sphere partition function \cite{CHM}. Of course, both of these coefficients are related to c-theorems in higher dimensions \cite{ctheorem,johnc,fthem}.

As shown in \cite{Mezei0}, the linear contribution in \req{defo} vanishes for a general CFT in arbitrary dimensions. The second order contribution was studied by Mezei in \cite{Mezei}. There he shows that for general holographic theories, this quadratic term is fully determined by the central charge $\ctt$ appearing in the two-point function of the stress tensor \cite{Osborn:1993cr}:
\beq
\braket{T_{\mu\nu}(x)\,T_{\rho\sigma}(0)}=\frac{\ctt}{x^{2d}}\,\mathcal{I}_{\mu\nu,\rho\sigma}(x)\,,
\eeq
where $\mathcal{I}_{\mu\nu,\rho\sigma}$ is a fixed dimensionless tensor.
In particular, $s_2^{\ssc (d)}$ and $\ctt$ are related through the compact expression \cite{Mezei}
\beq\label{meze}
s^{\ssc (d)}_{2}(V)=\ctt\, \frac{\pi^{\frac{d+2}{2}}(d-1)}{2^{d-2}\Gamma(d+2)\Gamma(d/2)}\sum_{\ell,m_1,...,m_{d-3}}a^2_{{\ell},m_1,...,m_{d-3}}
\frac{\Gamma(d+\ell-1)}{\Gamma(\ell-1)}\times\left\lbrace \begin{array}{cll}
 \pi/2 && d\text{ odd,}\\ 
1 && d\text{ even}.\end{array}\, \right.
\eeq

Of course, this result clearly resembles our general conjecture \reef{conj1} for $d=3$ CFTs: both $\sigma^{\ssc (3)}$ and $s^{\ssc (d)}_{2}$ are universal $\mathcal{O}(\epsilon^2)$ corrections to the EE of a smooth surface, (\eg $(\Omega-\pi/2)^2\sim \epsilon^2$ with $\epsilon\ll 1$ as $\Omega\rightarrow \pi/2$) and both coefficients are fully determined by $\ctt$. There are also some differences though: $\sigma^{\ssc (3)}$ characterizes a very particular deformation, namely the one which makes a sharp corner appear in the entangling surface, while $s^{\ssc (d)}_{2}$ encodes the contribution from a completely general smooth deformation of a hypersphere in general dimensions. On the other hand, as we have explained, the structure of EE divergences changes when the entangling surface $\partial V$ contains a conical singularity. In particular, the universal term for smooth surface is constant (logarithmically divergent) for odd (even) dimensional theories. However, that appearing with a conical singularity has a logarithmic (logarithmic$^2$) divergence for odd (even) $d$ --- compare \req{smooth} with \req{cone}. Therefore, if $s^{\ssc (d)}_{2}$ is to capture the corner contribution $\sigma^{\ssc (3)}$ and its natural extensions to higher-dimensions $\sigma^{\ssc (d)}$, the corresponding calculation must involve the appearance of an extra logarithmic divergence $\log (R/\delta)$ in each case.\footnote{Constructions involving the appearance of analogous logarithmic contributions can be found \eg in \cite{sing,igor7}.} We will see that this is indeed the case, and how by choosing particular deformations of $S^{d-2}$ which make infinitesimal conical singularities appear in the surface of the hypersphere, the corner coefficients $\sigma^{\ssc (d)}$ can be identified in general dimensions using eq.~\reef{meze}.  At the same time, this approach provides a general holographic proof of our conjecture \reef{conj1} and also extends it to higher dimensions as in eq.~\reef{CON}. We start below with the $d=3$ case, corresponding to our original conjecture \reef{conj1}. 


\subsection{Three dimensions}

\begin{figure}[h]
\center
\includegraphics[scale=0.6]{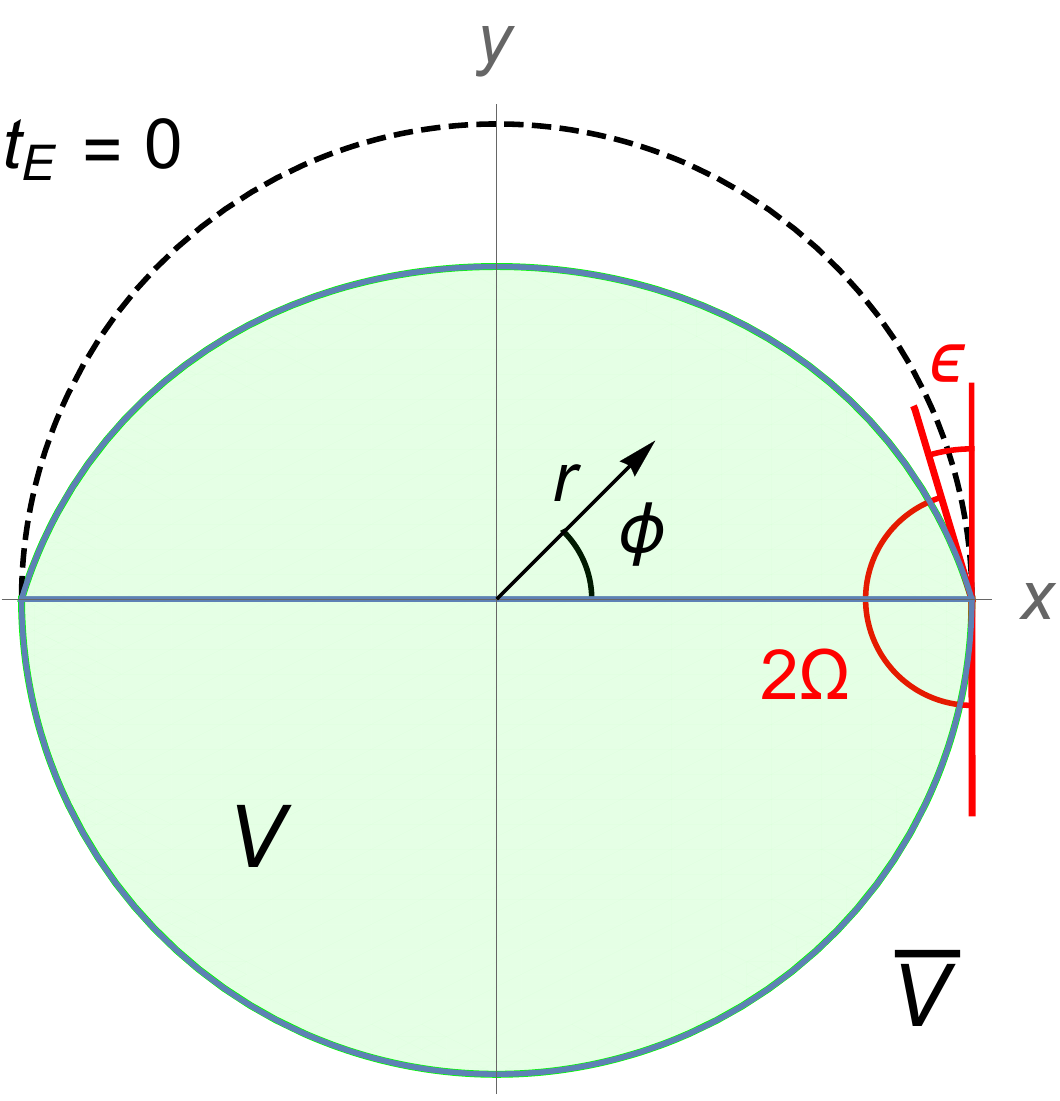}  \caption{An entangling region $V$ whose boundary corresponds to a deformed $S^1$ in eq.~\reef{rb1}. Two infinitesimal corner singularities appear at $\phi=0$ and $\pi$, with opening angle $2\Omega=\pi-\epsilon$.}    
\centering \label{Rb1}
\end{figure}

Consider calculating the EE in a three-dimensional CFT for a deformed circular entangling surface. In particular, we  consider the following deformation of the entangling surface 
\beq\labell{rb1}
r(\phi)/R=\left\lbrace \begin{array}{cll}
1-\epsilon \sin \phi&& \phi \in [0,\pi]\, ,\\ 
1 && \phi \in (\pi,2\pi)\, ,\end{array}\, \right.
\eeq
parametrized by the polar coordinate $\phi\in [0,2\pi)$ --- see Figure \ref{Rb1}. Below the $x$-axis, the circle remains undeformed but in the upper half plane, the surface is deformed in a way such that the surface is continuous for all values of $\phi$, but two infinitesimal corners or kinks are introduced at the $x$-axis, \ie at $\phi=0,\pi$. It is straightforward to check that the deficit angle at both of these points is given by $\epsilon$  In particular, we see from \req{rb1} that for $|\phi|\ll1$, the position of the entangling surface in Euclidean coordinates is given by 
\begin{align}
y&=\ \phi R + \mathcal{O}(R\phi^3)\, ,\\
x&=\left\lbrace 
\begin{array}{cll}
R-\epsilon\,R\,\phi+\mathcal{O}(R\phi^2) && \text{for}\ \phi\ge0\,,\\
R\ +\ \mathcal{O}(R\phi^2)&& \text{for}\ \phi\le0\,.\end{array}\, \right.
\end{align}
Hence, there is a kink in the entangling surface as it passes through the $x$-axis with $x\simeq R-\epsilon y$ for $y\ge0$ and $x\simeq R$ for $y\le0$. Further the slope of the tangent to the deformed surface for positive $\phi$ (with respect to the vertical) is precisely equal to $\epsilon$. This slope equals the tangent of the deficit angle and so we have: $\tan(\pi-2\Omega)\simeq (\pi-2\Omega)\simeq \epsilon$. Of course, a similar analysis of the second kink at $\phi=\pi$ yields the same deficit angle --- see Figure \ref{Rb1}.


%
Now our strategy is to `Fourier transform' the deformation of the circular profile in eq.~\reef{rb1} so that it may be represented as in eq.~\reef{mezei1}. We may then evaluate the universal contribution to the EE using eq.~\reef{meze}. For $d=3$, an orthonormal basis of real polar harmonics is given by
\beq
Y_{\ell}^{(c)}=\frac{1}{\sqrt{\pi}}\cos(\ell\phi)\, ,\quad Y_{\ell}^{(s)}=\frac{1}{\sqrt{\pi}}\sin(\ell\phi).
\eeq
With these functions, the corresponding coefficients in eq.~(\ref{mezei1}) become
\begin{align}
a_{\ell}^{(s)}&=-\frac{1}{\sqrt{\pi}}\int_0^{\pi}\sin\phi\, \sin({\ell}\phi)\,d\phi=-\left(\frac{\sin({\ell}\pi)}{\sqrt{\pi}(1-{\ell}^2)} \right)\, ,\\ \notag
a_{\ell}^{(c)}&=-\frac{1}{\sqrt{\pi}}\int_0^{\pi}\sin\phi\, \cos({\ell}\phi)\,d\phi=-\left(\frac{1+\cos({\ell}\pi)}{\sqrt{\pi}(1-{\ell}^2)}\, \right)\, .
\end{align}
However, the only non-vanishing components are:
\beq
a_1^{(s)}=-\frac{\sqrt{\pi}}{2}\qquad{\rm and} \qquad a_{2k}^{(c)}=\frac{2}{\sqrt{\pi}(4k^2-1)}\quad {\rm for}\ \ k=0,1,2,\cdots\,.
\eeq
Substituting these expressions into \req{meze}, we find
\beq \label{s3s}
s^{\ssc (3)}_{2}(V)=\frac{\pi^2\ctt}{2}\sum_{k=1}\frac{k}{4k^2-1} \, .
\eeq
For large $k$, the summand above is approximately $1/(4k)$ and so this sum is logarithmically divergent. If we regulate by cutting off the sum at $k=k_{\rm max}$, we find
\beq\labell{s3s1}
s^{\ssc (3)}_{2}(V)=\frac{\pi^2\ctt}{12}\,\Big[\,\gamma-1+
\log4+
\log (k_{\rm max})+\mathcal{O}(1/k_{\rm max}) \Big] \, ,
\eeq
where $\gamma$ is the Euler-Mascheroni constant. 

As discussed above, we anticipated the appearance of this new logarithmic divergence due to the presence of the two corner singularities in the deformed entangling surface \reef{rb1}. We can understand the divergent term physically as follows: We can think of $\ell$ in each Fourier mode as the physical (dimensionless) wavenumber of the corresponding perturbation of the circle. Hence, they can be associated with the corresponding wavelengths through $\lambda\sim 2\pi R/\ell$. Now, we should truncate the sum in \req{s3s} when these wavelengths are of the order of the UV cutoff, \ie when
$\ell\sim 2\pi R/\delta$. Hence, it is natural to set $\ell_{\rm max}=2k_{\rm max}=2\pi R/\delta$ in \req{s3s1}. This identification gives rise to exactly the desired logarithmic term required to match that appearing in the corner contribution to the EE. In particular, using eqs.~\reef{smooth} and \reef{defo}, as well as $\epsilon=\pi -2\Omega$, we find
\beq
S^{\rm univ}(V)= -\frac{\pi^2}{3}\,\ctt\,(\pi/2-\Omega)^2\, \log(R/\delta)\, .
\eeq
Given this expression and recalling that this term contains two corner contributions, we find the expected result\footnote{Mark Mezei found the same result independently using a similar procedure. We thank him for discussions on his calculations.}
\beq
\sigma^{\ssc (3)}=\frac{\pi^2}{24}\,\ctt\, .
\eeq

The elegant result \reef{meze} of \cite{Mezei} was the essential ingredient in our discussion above. The analysis in \cite{Mezei} involves the expression for holographic entanglement entropy constructed in \cite{Dong} for bulk theories of higher derivative gravity, whose Lagrangian does not contain derivatives of the
Riemann tensor. Hence the above discussion extends the proof of our original conjecture \reef{conj1} to this (infinite) class of general holographic theories. Of course, it is very likely that \req{meze} applies for all three-dimensional CFTs, in which case the present discussion would provide a general proof of \req{conj1}.

\subsection{General higher-dimensional case}

\begin{figure}[h]
\center
 \subfigure[]{\label{Rb3} \includegraphics[scale=.48]{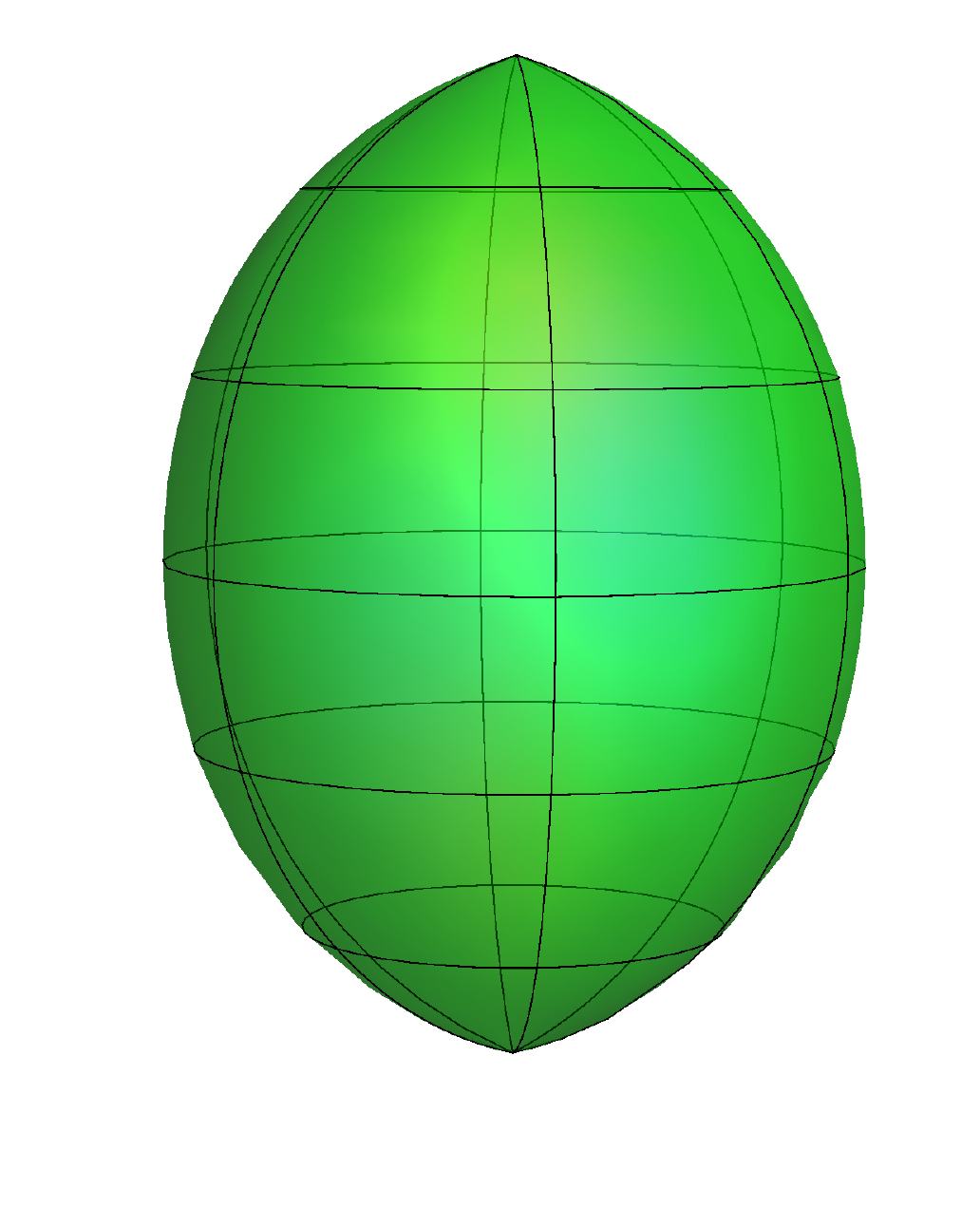}} \hspace{3cm}
 \subfigure[]{\label{Rb2}\includegraphics[scale=.54]{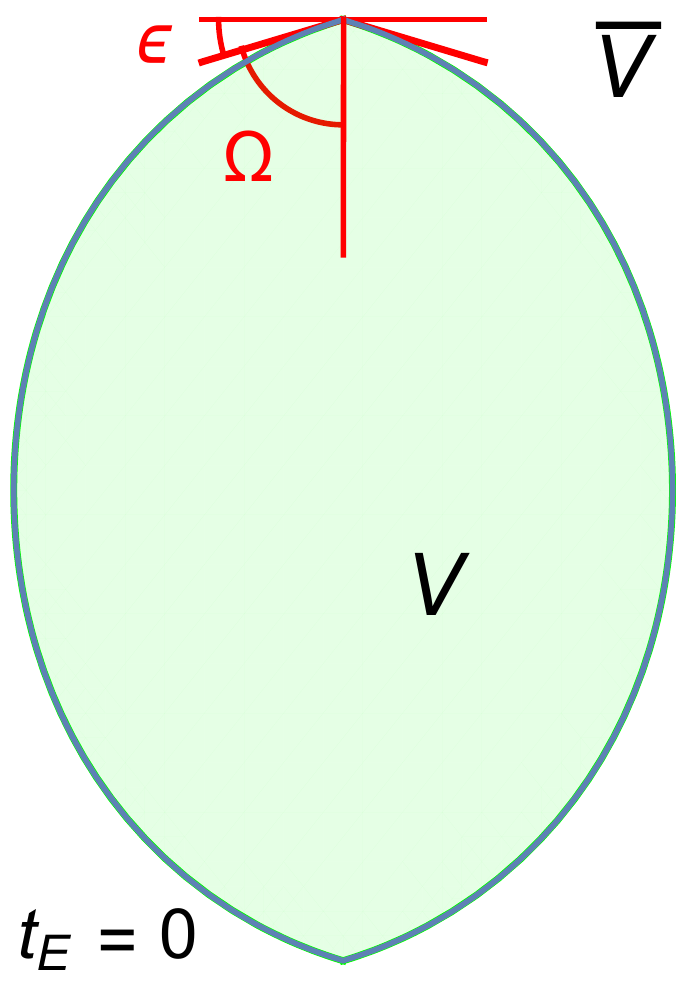}}   
\caption{In panel (a), we show a two-sphere deformed according to \req{defor} with $\epsilon=0.3$. We observe the appearance of two conical singularities in the poles. In panel (b), we plot a cross-section of the same surface. The deficit angle of the conical singularities is determined by $\epsilon=\pi/2-\Omega$ for small $\epsilon$.}    
\centering
\end{figure}

Let us now turn to the higher-dimensional case. In particular, we consider a spherical entangling surface in a $d$-dimensional CFT with $d\geq4$.  We then make the following infinitesimal deformation of the sphere $S^{d-2}$,
\beq\labell{defor}
r(\Omega_{d-2})/R=1-\epsilon \sin(\theta)\, ,
\eeq
where $\theta$ is the polar angle $\theta \in [0,\pi]$.
With this deformation, the $S^{d-2}$ looks like a $(d-2)$-dimensional rugby ball, which we denote $Rb^{d-2}$
--- see Figure \ref{Rb3}. Hence, there are two conical singularities at the poles $\theta=0$ and $\pi$. The deformation parameter $\epsilon$ in \req{defor} determines the deficit angles at these singularities through $\epsilon=\pi/2-\Omega$ --- see Figure \ref{Rb2}.\footnote{This relation is slightly different from our construction for $d=3$ in the previous section.} 

Now our analyis is simplified since $Rb^{d-2}$ preserves an SO($d-2$) subgroup of the SO($d-1$) group of isometries of the round $S^{d-2}$. The Laplace operator on $S^{d-2}$ can be written recursively as\footnote{We use the following set of angular coordinates: $\theta,\theta_2,...,\theta_{d-3}\in[0,\pi]$ and  $\phi\in[0,2\pi)$.} 
\beq
\triangle_{S^{d-2}}=\frac{1}{(\sin\theta)^{d-3}}\frac{\partial}{\partial\theta}\left[(\sin\theta)^{d-3}\frac{\partial}{\partial\theta} \right]+\frac{1}{\sin^2\theta}\triangle_{S^{d-3}}\, ,
\eeq
where we have used the label $\theta$ for the first polar coordinate in each case. 
To describe the $Rb^{d-2}$ hypersurface, we need only consider the subset of hyperspherical harmonics, $Y_{\ell}(\theta)$, depending only on $\theta$, \ie we concentrate on $m_1=\cdots=m_{d-3}=0$ from the general hyper-spherical harmonics. In this case, the Laplace equation \reef{hypers} simplifies to
\beq\labell{hypp}
\frac{1}{(\sin\theta)^{d-3}}\frac{\partial}{\partial\theta}\left[(\sin\theta)^{d-3}\frac{\partial}{\partial\theta} \right]Y_{\ell}(\theta)=-\ell(\ell+d-3)\,Y_{\ell}(\theta)\,.
\eeq
The general solution to \req{hypp} is given by
\beq\label{yh}
Y_{\ell}(\theta)=\frac{p_{\ell}}{\sin^{(d-4)/2}\theta}\,P^{(d-4)/2}_{\ell+(d-4)/2}(\cos\theta)+\frac{q_{\ell}}{\sin^{(d-4)/2}\theta}\,Q^{(d-4)/2}_{\ell+(d-4)/2}(\cos\theta)\, ,
\eeq
where $P^{m}_{\ell}(x)$ and $Q^{m}_{\ell}(x)$ are associated Legendre polynomials (ALPs) of the first and second kind, respectively, and where $p_{\ell}$ and $q_{\ell}$ are normalization constants, which will be fixed by demanding $\int d\Omega_{d-2}\,Y_{\ell}(\theta)\,Y_{\ell^{\prime}}(\theta)=\delta_{\ell\ell^{\prime}}$.
 In even dimensions, the ALPs of the second kind are generically non-normalizable and so we set $q_\ell=0$ in this case.\footnote{For example, for $d=4$, we will have: $Y_{\ell}(\theta)=\,\sqrt{\frac{(2\ell+1)}{4\pi}} P_{\ell}(\cos \theta)$, which are the usual zonal harmonics on $S^2$.} Similarly, in odd dimensions, the ALPs of the first kind that are non-normalizable and we must set $p_\ell=0$. Hence we find:
\beqa
d\ {\rm even}:&&\qquad p_{\ell}= \hat{p}_{\ell}
\,,\qquad\quad q_{\ell}=0\,;\\
\nonumber
d\ {\rm odd}:&&\qquad
p_\ell=0\,, \qquad \quad\  q_{\ell}=\frac{2}{\pi}\,\hat{p}_{\ell} \,.
\eeqa
where 
\beq
\hat{p}_{\ell}^2 \equiv\frac{\Gamma[\ell +1]}{ \Gamma [d+\ell-3]}\frac{(d+2 \ell-3) }{2\Omega_{d-3}}
\qquad{\rm with}\qquad
\Omega_{d-3}=\frac{2\pi^{(d-2)/2}}{\Gamma[(d-2)/2]}\,,
\eeq
which is the volume of a unit $S^{d-3}$. 

Hence with the deformation in \req{defor}, we evaluate the coefficients in the expansion \reef{mezei1} as 
\beqa \label{aev}
a_{\ell}&=&q_{\ell}\ \sqrt{\Omega_{d-3}}\ \int_0^{\pi}d\theta\ \sin^{d/2}\!\theta\ Q_{\ell+(d-4)/2}^{(d-4)/2}(\cos\theta)
\,,\qquad d\ {\rm odd}\, ;\\
\nonumber
&=&p_{\ell}\ \sqrt{\Omega_{d-3}}\ \int_0^{\pi}d\theta\ \sin^{d/2}\!\theta\ P_{\ell+(d-4)/2}^{(d-4)/2}(\cos\theta)\,,\qquad d\ {\rm even}\, .
\eeqa
Now for both odd and even dimensions, we find that $a_{\ell}^2$ are only nonvanishing for even values of $\ell$, \ie $\ell=2k$. Interestingly, the final value can be expressed in a closed form valid for all dimensions $d\geq 4$, odd or even, which reads
\beq
a_{2k}^2=\frac{(4k+d-3)\,\pi^{d/2-2}\,\Gamma\left[\frac{d-1}{2}\right]^2\Gamma\left[k-\frac{1}{2}\right]^2\Gamma\left[k+\frac{1}{2}\right]\Gamma\left[k+\frac{d-3}{2}\right]}{4\,\Gamma\left[\frac{d-2}{2}\right]\Gamma\left[k+1\right]\Gamma\left[k+\frac{d}{2}-1\right]\Gamma\left[k+\frac{d}{2}\right]^2}\, .
\eeq
We can now use this expression in Mezei's formula \reef{meze} to find
\beq\label{harvey}
s^{\ssc (d)}_{2}(V)=\ctt \frac{\pi^{\frac{d+2}{2}}(d-1)}{2^{d-2}\Gamma(d+2)\Gamma(d/2)}\sum_{k}A_k\times \left\lbrace \begin{array}{cll}
 \pi/2&& d\text{ odd}\, ,\\ 
1 && d\text{ even}\, ,\end{array}\, \right.
\eeq
where
\beq
A_k=\frac{(4k+d-3)\,\pi^{d/2-2}\,\Gamma\left[\frac{d-1}{2}\right]^2\Gamma\left[k-\frac{1}{2}\right]^2\Gamma\left[k+\frac{1}{2}\right]\Gamma\left[k+\frac{d-3}{2}\right]}{4\,\Gamma\left[\frac{d-2}{2}\right]\Gamma\left[k+1\right]\Gamma\left[k+\frac{d}{2}-1\right]\Gamma\left[k+\frac{d}{2}\right]^2}
\,\frac{\Gamma(2k+d-1)}{\Gamma(2k-1)}\, .
\eeq
Now the behaviour of $A_k$ as $k\rightarrow \infty$ is given by
\beq
A_k= \frac{2^d\pi^{d/2-2}\Gamma[\frac{d-1}{2}]^2}{\Gamma[\frac{d-2}{2}]}\frac{1}{k}+\mathcal{O}(1/k^2)\, ,
\eeq
and so \req{harvey} becomes
\beq \labell{s2c}
s^{\ssc (d)}_{2}(V)=\ctt \frac{\pi^{\frac{d+2}{2}}(d-1)}{2^{d-2}\Gamma(d+2)\Gamma(d/2)}\left[\frac{2^d\pi^{d/2-2}\Gamma[\frac{d-1}{2}]^2}{\Gamma[\frac{d-2}{2}]}\,\log(k_{\rm max})+\cdots \right]\times \left\lbrace \begin{array}{cll}
 \pi/2&& d\text{ odd}\, ,\\ 
1 && d\text{ even}\, ,\end{array}\, \right.
\,
\eeq
where again we have introduced a cut-off $k_{\rm max}$, and the ellipsis refers  to terms which do not scale with $k_{\rm max}$ (or with inverse powers of this cut-off). Hence we can see that the two conical singularities at the poles of the sphere again give rise to an additional logarithmic divergence in the EE. Comparing to \req{smooth},  the universal contribution to the EE becomes
\begin{eqnarray}\labell{smoothi}
S^{\rm univ}(V)= \ctt\,\frac{2(d-1)(d-2)\pi^{d-1}\,\Gamma[\frac{d-1}{2}]^2}{\Gamma[\frac{d}{2}]^2\,\Gamma[d+2]}\,\epsilon^2 \times\left\lbrace \begin{array}{cll}
\hspace{-1cm}(-1)^{\frac{d-1}{2}} \,\pi/2\,\log(k_{\rm max})&& d\quad \text{odd}\, ,\\ 
 (-1)^{\frac{d-2}{2}}\,  \,\log(k_{\rm max}) \log (R/\delta)&& d\quad \text{even}\,
.\end{array}\, \right.
\end{eqnarray}

This expression above contains contributions from two conical singularities (one at each pole of the sphere) and so we must divide by $2$ in order to extract $\sigma^{\ssc (d)}$  in all cases. In odd dimensions, we  replace $\log(k_{\rm max})= \log (\pi R/\delta)$, just like for $d=3$, which allows us to identify 
 \beq
\sigma^{\ssc (d)}=\ctt\,\frac{(d-1)(d-2)\pi^{d}\,\Gamma[\frac{d-1}{2}]^2}{8\,\Gamma[\frac{d}{2}]^2\,\Gamma[d+2]}\, , \quad  d\ \ \text{ odd}\,.
\eeq
We show some explicit values in Table \ref{tbl1}. Note that this formula for general odd $d$ properly incorporates the result \reef{conj1} for $d=3$. Interestingly, the analytic values given here for $d\ge5$ are not easily accessible through the standard calculation using the Ryu-Takayanagi (RT) prescription \cite{rt0} for an entangling surface containing a conical singularity. In particular, the corresponding corner functions $a^{\ssc (5,7,\cdots)}(\Omega)$ of the opening angle are given by complicated implicit expressions which can only be treated numerically, \eg see \cite{sing} for a discussion of the $d=5$ case. 

In even dimensions, there is a subtlety in the identification of $\log(k_{\rm max})$ with $\log (R/\delta)$, which requires comment below. Indeed, in this case, the correct substitution is $\log(k_{\rm max})=\frac12\,\log(\pi R/\delta)$ instead. Taking this into account, we find
 \beq
\sigma^{\ssc (d)}=\ctt\,\frac{(d-1)(d-2)\pi^{d-1}\,\Gamma[\frac{d-1}{2}]^2}{8\,\Gamma[\frac{d}{2}]^2\,\Gamma[d+2]}\, , \quad  d\ \ \text{ even}\,.
\eeq
We have explicitly verified that this formula exactly reproduces the results obtained using the RT prescription for a cone of opening angle $\Omega=\pi/2-\epsilon$ for $d=4, 6, 8, 10, 12, 14$ --- see Table \ref{tbl1}.

Let us now comment on the reason behind the factor $1/2$ which appears above in the even-dimensional case when relating the highest wavenumber $k_{\rm max}$ to the short distance cut-off $\delta$.\footnote{The very same factor $1/2$ was observed to appear in \cite{sing,igor7} and \cite{ben2} when computing $a^{(d)}(\Omega)$ for $d=4$ and $6$, respectively.} An illustrative way of understanding this factor consists of comparing our calculation here with the one performed using the RT prescription for an entangling region consisting of a cone (\eg in $d=4$). In the latter, the $\log(R/\delta)^2$ term arises from an integral of the form $\int_R^{\delta} \frac{dr}{r}\log(r/\delta)=\frac12\, \log(R/\delta)^2$, \eg see \cite{sing}. In our present calculation, the two logarithms arise separately. One is a fixed overall factor in the universal term \reef{smooth} of a generic smooth surface in even dimensions. The other is obtained  from the sum $\sum_{k=1}^{k_{\rm max}} 1/k$, which produces the $\log(k_{\rm max})$ factor. The latter is like performing the integration $\int_R^{\delta} \frac{dr}{r}$ separate from the logarithmic factor and we observe that the naive substitution $\log(k_{\rm max})\rightarrow \log(R/\delta)$ fails to yield the correct answer by precisely a factor $2$ \cite{igor7}. 
This holographic calculation suggests that there should not be a fixed infrared scale in the overall logarithmic factor appearing in the universal term \reef{smooth}. At least for the perturbations at smaller wavelengths, this IR scale should match the wavelength of the perturbation. More pragmatically, in order to produce the correct $\frac12\,\log(R/\delta)^2$ factor in our calculation, we need to make the replacement  $\log(k_{\rm max})\rightarrow \frac12\, \log(R/\delta)$ instead. From this discussion, it is also clear that the subtlety is exclusive to theories in even dimensions since in odd dimensions, the analogous integral in the RT calculation is of the form $\int_R^{\delta} \frac{dr}{r}=\log(R/\delta)$, and therefore it does not produce any additional factor with respect to our naive substitution $\log(k_{\rm max})\rightarrow \log( R/\delta)$.\footnote{We omitted the factor $\pi$ in the relation $k_{\rm max}\sim \pi R/\delta$ in this paragraph as it does not play any role in the final result and it might however be confusing.} In particular, we have seen that the $d=3$ calculation gives rise to the correct $\sigma^{\ssc (3)}/\ctt$ ratio.

With this explanation we conclude our proof of the relation $\sigma^{\ssc (d)}/\ctt$ in general dimensions for general holographic theories, which is summarized in \req{CON}. Again, if \req{meze} holds for general CFTs in arbitrary dimensions, our results here would provide a general proof of \req{CON}. In any event, we conjecture that \req{CON} applies not only for holographic CFTs but for general CFTs.


\section{Cone coefficients for R\'enyi entropy} \labell{renyi}

A conical defect in an otherwise smooth entangling surface will introduce a new universal contributions to the \ren entropy \reef{cone}, analogous to those discussed above for the EE. In particular, the corresponding \ren cone contributions $a^{\ssc (d)}_n(\Omega)$ behave as in eq.~\reef{limi} for large opening angles. In \cite{bueno3}, we considered the corresponding corner coefficient $\sigma^{\ssc (3)}_n$ controlling this universal contribution to the \ren entropy for an almost smooth entangling surface. There, we argued that our original conjecture \reef{conj1} is a particular case of a more general relation connecting $\sigma_n^{\ssc (3)}$ with the scaling dimension of the corresponding twist operators $h_n$ with
\begin{eqnarray}\labell{conj3}
\sigma_n^{\ssc (3)}=\frac{1}{\pi}\,\frac{h_n}{n-1}\, .
\end{eqnarray}
In \cite{bueno3}, we verified that \req{conj3} is satisfied for all integer values of $n$ and in the limit $n\rightarrow \infty$ both for a free scalar and a free fermion --- see also \cite{dowker}. 
 
While we provide a precise definition of the scaling dimension $h_n$ in appendix \ref{Atwist}, a key result for our present purposes will be \cite{holoren,twist}\footnote{Ref.~\cite{Perlmutter:2013gua} considers an analogous derivative of the \ren entropy.}
\begin{eqnarray}\labell{twist}
\left. \partial_n h_n \right\vert_{n=1}=2\pi^{\frac{d+2}{2}}\frac{\Gamma[{d}/{2}]}{\Gamma[d+2]}\ \ctt\, .
\end{eqnarray}
That is, that the first derivative of the scaling dimension at $n=1$ is determined by the central charge $\ctt$. Recalling that $h_1=0$, we may use eq.~\reef{twist} to write the leading term in an expansion about $n=1$ as,
\beq
h_n \ \stackrel{n\to1}{=}\  2\pi^{\frac{d+2}{2}}\frac{\Gamma[{d}/{2}]}{\Gamma[d+2]}\ \ctt \ (n-1)\ +\  \cO\left((n-1)^2\right)\,.
\labell{expand}
\eeq
In particular then, for $d=3$, we have $h_n \simeq \frac{\pi^3}{24}\,\ctt \, (n-1)$ and upon substituting this expansion into \req{conj3}, we see that it reduces 
to our original conjecture \reef{conj1} for the EE at $n=1$. Therefore, the new conjecture \reef{conj3} is supported by all of the evidence supporting eq.~\reef{conj1}, including calculations for free scalars and fermions \cite{bueno1,bueno2,Elvang,dowker}, as well as the general proof for holographic theories given above --- see also \cite{rxm}.

Now these results suggest a natural extension to the \ren cone coefficients in higher dimensions. In particular, the expansion \reef{expand} of the scaling dimension in general dimensions suggests that  our result \reef{CON} extends to
 \begin{eqnarray}\labell{new9}
\sigma_n^{\ssc(d)}=\frac{h_n}{n-1}\ \frac{(d-1)(d-2)\,\pi^{\frac{d-4}{2}}\,\Gamma\left[\frac{d-1}2\right]^2}{16\ \Gamma[{d}/{2}]^3}\ \times\,\left\lbrace 
\begin{array}{cll}
\pi && d\,\, \text{odd}\,,\\
1&& d\text{ even}\,.\end{array}\, \right.
\end{eqnarray}
Hence for the next few dimensions, eq.~\reef{conj3} is supplemented by
\beq\labell{expl}
\sigma_n^{\ssc(4)}=\frac{3\pi}{32}\,\frac{h_n}{n-1}\,,\qquad
\sigma_n^{\ssc(5)}=\frac{16}{9}\,\frac{h_n}{n-1}\,,\qquad
\sigma_n^{\ssc (6)}=\frac{45\pi^2}{512}\,\frac{h_n}{n-1}\,.
\eeq
Now the scaling dimension of twist operators in the free scalar or free fermion theories can be straightforwardly calculated using heat kernel techniques in any number of dimensions \cite{twist} --- see also \cite{dowker}. Hence it would be interesting if one could directly evaluate the coefficients $\sigma_n^{\ssc(d)}$ for these theories to provide further evidence supporting \req{new9}. In the following, we take some steps in this direction for the cases of $d=4$ and $d=6$.

\subsection{Four dimensions} \labell{four}

The universal contribution to the \ren entropy of a CFT for a general region can be expressed in terms of a geometric integral over the entangling surface \cite{ben1}\footnote{In this expression, $h_{ab}$ is the induced metric on the two-dimensional entangling surface $\partial V$ and ${\cal R}$ is the corresponding (intrinsic) Ricci scalar. The extrinsic curvature is denoted by $K^i_{ab}$ where $a,b$ and $i$ denote the two tangent directions and the two transverse directions to $\partial V$, respectively. Hence $\tr K^2\equiv K^i_a{}^{\,b}K^i_b{}^{\,a}$ and $K^2\equiv K^i_a{}^{\,a} K^i_b{}^{\,b}$ where the indices are raised with the inverse metric $h^{ab}$. Further, $C^{ab}{}_{ab}$ is the background Weyl curvature also traced with $h^{ab}$.}
\beq
S^{ \rm univ}_n=-\frac{\log\left(R/\delta\right)}{2\pi}\,\int_{\partial V} d^2y\sqrt{h}\left[ f_a(n)\,{\cal R} + f_b(n)\left( \tr K^2-\frac12 K^2\right)-f_c(n)\, C_{ab}{}^{ab}
\right]
\labell{game}
\eeq
where the functions $f_{a,b,c}(n)$ are independent of the geometry of the entangling surface or the background spacetime. In the limit $n\to1$, these functions are related to the coefficients appearing in the trace anomaly:\footnote{Note that we have adopted a convention where for a massless free real scalar field these coefficients are given by: $a=1/360$ and $c=1/120$. This may be contrasted with the conventions of \cite{ben1} where the corresponding coefficients are given by: $a=1=c$.}
\beq
f_a(n=1)=a\,,\qquad f_b(n=1)=c=f_c(n=1)\,,
\labell{game3}
\eeq
and with these values, eq.~\reef{game} reduces to the corresponding expression for the universal contribution to the EE in four-dimensional CFTs \cite{solo9}. 

Further, two of these coefficients are readily calculated for a massless free real scalar or a massless free Dirac fermion \cite{ben1,igor8,Fursaev:2012mp,Casini:2010kt,Fursaev:1993hm,DeNardo:1996kp}:
\beqa
f^{\rm scalar}_a(n)&=&\frac{(1+n)(1+n^2)}{1440n^3}\,,\qquad\quad\ \,
f^{\rm scalar}_c(n)\ =\ \frac{(1+n)(1+n^2)}{480n^3}\,,
\labell{fabc}\\
f^{\rm fermion}_a(n)&=&\frac{(1+n)(7+37n^2)}{2880n^3}\,,\qquad
f^{\rm fermion}_c(n)\ =\ \frac{(1+n)(7+17n^2)}{960n^3}\,.
\nonumber
\eeqa
Ref.~\cite{ben1} conjectured that the following relation held for all four-dimensional CFTs
\beq
f_b(n)=f_c(n)\,,
\labell{conjx9}
\eeq
and they provided numerical evidence for the free scalars and fermions that these two coefficients were identical. Further support was provided by \cite{aitor9}, which argued that it also held for free Maxwell fields, ${\cal N}=4$ super-Yang-Mills and a broad class of holographic CFTs. Ref.~\cite{aitor9} also argued that in general, $f_a(n)$ and $f_c(n)$ are related with \beq
f_c(n)=n\,\left( \frac{a - f_a(n)}{n-1}-\partial_n f_a(n)\right)\,.
\labell{aitor8}
\eeq

Now we may apply these results to determine the universal contribution to the \ren entropy coming from a conical entangling surface in $d=4$.\footnote{We closely follow the analogous derivation for the entanglement entropy, \ie $n=1$, in \cite{igor7}.} Parametrizing the cone in spherical coordinates $(t_E,r,\theta,\phi)$ as $t_E=0$, $\theta=\Omega$ --- see Figure \ref{con} --- it is easy to find the two normal vectors $n^1=\partial_{t_E}$, $n^2=r\partial_\theta$. The only non-vanishing component of the extrinsic curvatures associated to these vectors is $K_{\phi\phi}^2=1/2\, r \sin 2\Omega$. Using this result in \req{game}, one finds that the only nonvanishing contribution to the universal term \reef{game} comes from the term proportional to $f_b(n)$
\beq
S^{ \rm univ}_n=-\frac{1}{2}\,f_b(n)\,\frac{\cos^2\Omega}{\sin{\Omega}}\log\left(R/\delta\right)\,\int_{r_{\rm min}}^{r_{\rm  max}}   \frac{dr}{r}\, ,
\labell{game1}
\eeq
where we have introduced UV and IR cut-offs in the radial integral, $r_{\rm min}$ and $r_{\rm max}$, respectively. As discussed in the previous section, the naive replacement $\log (r_{\rm max}/r_{\rm min})\rightarrow \log (R/\delta)$ fails to give the right answer by a factor $2$. Taking this into account, we find the correct result is
\beq
S^{\rm univ}_n = -\frac{1}{4}\,f_b(n)\, \frac{\cos^2\Omega}{\sin{\Omega}}\, \log^2(R/\delta)\,.
\labell{result}
\eeq
Note that this calculation has fixed the entire angular dependence of the cone contribution with
\beq
a_n^{\ssc (4)}(\Omega)=\frac{1}{4}\,f_b(n)\, \frac{\cos^2\Omega}{\sin{\Omega}}\,.
\labell{result1}
\eeq
Of course, this function exhibits the appropriate behaviour in the limit $\Omega\to\pi/2$ given in eq.~\reef{limi}. In particular, we have
\beq
\sigma_n^{\ssc (4)}= \frac{1}{16}\,f_b(n)\,.
\labell{result2}
\eeq
Finally combining the above expression with our conjecture, we arrive at
\beq
f_b(n)=  \frac{3\pi}2\,\frac{h_n}{n-1}\,.
\labell{result3}
\eeq

Now, the scaling dimension $h_n$ has been evaluated using heat kernel techniques for the free real scalar or free Dirac fermion theories in four dimensions as \cite{twist}
\beqa
h_n^{\rm scalar}&=&\frac{1}{720\pi} \frac{n^4-1}{n^3}\,,
\labell{globe}\\
h_n^{\rm fermion}&=& \frac{1}{1440\pi} \frac{(n^2-1)(7+17n^2)}{n^3}\ \,,
\nonumber
\eeqa 
which combined with eq.~\reef{result3} yields
\beqa
f_b^{\rm scalar}(n)&=& \frac{(1+n)(1+n^2)}{480n^3}\,,
\labell{globe}\\
f_b^{\rm fermion}(n)&=&   \frac{(1+n)(7+17n^2)}{960n^3}\ \,.
\nonumber
\eeqa 
For these free CFTs, we know that eq.~\reef{conjx9} certainly applies and hence we can compare these results to the expressions for $f_c(n)$ in eq.~\reef{fabc}. We find complete agreement with those expressions and hence we have additional support for our conjecture \reef{new9}.

We should note that by independent calculations, ref.~\cite{aitor9} derived
\beq
f_c(n)=  \frac{3\pi}2\,\frac{h_n}{n-1}\,,
\labell{result4}
\eeq
as a general result for four-dimensional CFTs. This result then connects the two conjectures in eqs.~\reef{new9} and \reef{conjx9}. That is, finding a general proof of the four-dimensional version of our conjecture for the \ren cone coefficient will provide a general proof of \req{conjx9} and vice versa. More generally if we accept both eqs.~\reef{conjx9} and \reef{aitor8} for general four-dimensional CFTs, our calculations here indicate that all three of the coefficients in eq.~\reef{game} are completely determined by the scaling dimension of the twist operator (as well as the A-type trace anomaly coefficient). In particular, we find \cite{aitor9}
\beqa\label{fabc1}
&& \partial_n\Big[(n-1)\,f_a(n)\Big]=\,a-\frac{3\pi}{2n}\,h_n\,.
\labell{total}\\
&&f_b(n)=f_c(n)= \frac{3\pi}2\,\frac{h_n}{n-1}\,.
\nonumber
\eeqa

As an example, eq.~\reef{fabc1} can be used to predict these coefficients for strongly coupled holographic CFTs dual to Einstein gravity \cite{aitor9}. In particular, we begin with the AdS/CFT correspondence in its simplest setting, where it describes a four-dimensional boundary CFT in terms of five-dimensional Einstein gravity in the bulk with the action   
\begin{align}
I=\frac{1}{16\pi G}\int d^5x\sqrt{g}\left[\frac{12}{L^2}+\mathcal R \right]\, ,
\labell{einst}
\end{align}
where $G$ is the four-dimensional Newton's constant, $L$ is the AdS$_{5}$ radius, and $\mathcal R$ is the Ricci scalar.  
In order to obtain $h_n^{\rm hol}$, we need to consider the thermal ensemble of the boundary CFT on the hyperbolic geometry appearing in the construction of \cite{CHM}, which is then equivalent to a topological black hole with a hyperbolic horizon. We refer the interested reader to \cite{holoren} for the detailed calculations and simply quote the result here:
\begin{align} 
h^{\rm hol}_n=\frac{L^{3}}{G}\ \frac{8n^4-4n^2-1-\sqrt{1+8n^2}}{256\,n^3}\, .\labell{laugh}
\end{align}
Of course, we also need the central charge $a$ for the boundary CFT \cite{whole}: $a^{\rm hol}=\frac{\pi\,L^3}{8\, G}$. Now, using the identities in \req{fabc1}, we can easily use the above expression to compute the $f_{a,b,c}(n)$ coefficients. We find 
\beqa
&& f^{\rm hol}_a(n)=\frac{\pi  L^3}{8 G}-\frac{\pi  L^3}{128\, G}\,\frac1{n^3} \left(6n^3-10n^2-n-1+\frac{2(n+1)(8n^2+1)}{3+\sqrt{8
   n^2+1}}\right)\,\,,
\nonumber\\\labell{total}
&&f^{\rm hol}_b(n)=f^{\rm hol}_c(n)= \frac{3 \pi \, L^3}{128\, G}\,\frac{n+1}{n^3} \left(2n^2+1-\frac{2} {3+\sqrt{8 n^2+1}}\right)\,\,.
\eeqa
From these expressions, it is not difficult to verify that $f^{\rm hol}_a(n=1)=a^{\rm hol}=\pi L^3/(8G)$, $f^{\rm hol}_b(n=1) = f^{\rm hol}_c(n=1)=c^{\rm hol}=\pi L^3/(8G)$, as expected.

\subsection{Six dimensions}
As we pointed out in the previous section, $a^{\ssc (4)}(\Omega)$ is proportional to $\ctt$ on general grounds for all values of $\Omega$ \cite{igor7}. However, this behaviour seems to be particular for four-dimensional CFTs (and, of course, also trivially for $d=2$ theories). For example, for six-dimensional holographic CFTs which are dual to Gauss-Bonnet gravity in the bulk, one finds \cite{sing} 
\begin{align}\labell{track2}
a^{\ssc (6)}(\Omega)=\frac{3}{1024}\, \frac{\cos^2\Omega}{\sin\Omega}\left(\left(\frac{5\pi^6}{56}\,\ctt+3A\right)-\left(\frac{\pi^6}{168}\,\ctt-3A\right)\cos2\Omega \right)\, ,
\end{align}
where $A$ is the universal coefficient of the A-type trace anomaly, \ie the coefficient appearing in the universal EE of a spherical entangling surface, as in \req{spa}. However, in the limit $\Omega\rightarrow \pi/2$, the dependence on $A$ cancels in the leading contribution \reef{limi} and one finds $\sigma^{\ssc (6)}/\ctt=\pi^6/14336$, as expected from \req{CON}. 
 
Interestingly, using the results in \cite{ben2,aitor9}, we can readily extend the validity of eqs.~\reef{CON} and \reef{new9a} for $d=6$ to a class of CFTs which extends beyond holographic theories. The idea is as follows: The universal term in the R\'enyi entropy for a general six-dimensional CFT can be written \cite{ben2}, in a analogous way to \req{game1} for $d=4$, with a sum of terms involving various combinations of intrinsic and extrinsic curvature tensors integrated over the entangling surface $\partial V$. For the EE (\ie $n=1$), each of these combinations is weighted by one of the Weyl anomaly coefficients $A$, $B_1$, $B_2$, $B_3$
\begin{align}\labell{track}
\braket{T^{\mu}{}_{\mu}}=\sum_{i=1}^3\,B_i\, I_i+2\,A\, E_6\, ,
\end{align} 
where $E_6$ is the Euler density of six-dimensional manifolds and the $I_i$ are independent invariants consisting of various contractions of the Weyl tensor \cite{Bastianelli:2000hi,check}. In \cite{ben2}, it was shown that for all theories satisfying 
\begin{align}\labell{cons6}
3B_3=B_2-\frac{B_1}{2}\, ,
\end{align}
the corresponding universal term (in a flat background) can be written as
\begin{align}\labell{univ6}
S^{\rm univ}=\log(R/\delta)\int_{\partial V} d^4y \sqrt{h}\left[2A\, E_4 +6\pi \left(B_2-\frac{B_1}{4} \right)J+B_3\, T_3 \right]\, ,
\end{align}
 where $E_4$ is the four-dimensional Euler density, while $J$ and $T_3$ are certain complicated combinations of extrinsic curvatures.\footnote{In particular, they are given by$J\equiv \frac{5K^2}{4}\left(\frac{K^2}{8}-\tr K^2\right)+(\tr K^2)^2+2\left(K\tr K^3- \tr K^4 \right)$ and $T_3\equiv (\nabla_a K)^2-\frac{25}{16}K^4+11K^2\tr K^2-6 (\tr K^2)^2-16 K \tr K^3+12 \tr K^4$ respectively.}
In theories satisfying \req{cons6},  \req{univ6} can then be used to determine $\sigma^{\ssc (6)}$ along the lines of the previous discussion for $d=4$. The final result is \cite{ben2}
\begin{align}\labell{sig6}
\sigma^{\ssc (6)}=\frac{27 \pi^3}{2}\, B_3\, .
\end{align}
Further, in \cite{aitor9}, it was proven that $B_3$ and $\ctt$ are related for general theories through
 \begin{align}\labell{track3}
 B_3=\frac{\pi^3}{193536}\,\ctt\, .
 \end{align}
 Combining these two results, one then finds $\sigma^{\ssc (6)}/\ctt=\pi^6/14336$, in agreement with our general formula \req{CON}. 
 
Let us add that the holographic CFTs dual to Gauss-Bonnet gravity satisfy the constraint in eq.~\reef{cons6} \cite{ben2}. Hence eq.~\reef{track2} provides the universal contribution from a conical entangling surface in $d=6$ for all angles in this class of theories. That is,
when we apply eq.~\reef{univ6} to evaluate the universal contribution from a conical entangling surface in $d=6$ and we use eq.~\reef{track3} to replace $B_3$ with $\ctt$, the final result matches that in eq.~\reef{track2}.
 
 Hence, we find that all six-dimensional CFTs satisfying \req{cons6} respect our new formula \reef{CON}. As shown in \cite{ben2}, these include at least some holographic theories like Lovelock gravity, but also other theories, such as the interacting $\mathcal{N} = (2, 0)$ theory describing a large number of coincident M5-branes, or a free $\mathcal{N} = (1, 0)$ hypermultiplet consisting of one Weyl fermion and 4 real scalars. Interestingly, this is the minimal free model for which \req{cons6} is fulfilled, \eg a single free scalar or a free fermion do not satisfy \req{cons6}. Indeed, the anomaly coefficients for a free real scalar and a free Dirac fermion read \cite{Bastianelli:2000hi}
 \beqa
B_1^{\rm scalar}&=&-\frac{1}{(4\pi)^3\, 540}\,,\qquad\quad\ \,
B_1^{\rm fermion}\ =\ -\frac{8}{(4\pi)^3\,135}\,,
\labell{b123}\\
B_2^{\rm scalar}&=&\frac{1}{(4\pi)^3\, 3024}\,,\qquad\quad\ \,\,\,\,\,
B_2^{\rm fermion}\ =\ -\frac{2}{(4\pi)^3\,315}\,, \\
B_3^{\rm scalar}&=&\frac{1}{(4\pi)^3\,2520}\,,\qquad\quad\ \,\,\,\,\,\,\,\,
B_3^{\rm fermion}\ =\ \frac{1}{(4\pi)^3\,126}\,.
\nonumber
\eeqa
Hence the constraint \req{cons6} is only satisfied when the theory contains 4 real scalars for each Weyl fermion, as in the $\mathcal{N}=(1,0)$ hypermultiplet. As a result, we do not yet have a proof of our general conjecture \req{CON} for free fields in $d=6$. 

The previous discussion can be extended to general values of the R\'enyi index $n$. In that case, the coefficients $B_i$ are replaced by certain functions $f_{B_i}(n)$ satisfying $f_{B_i}(n=1)=B_i$. In \cite{aitor9}, it was shown that $f_{B_3}(n)$ and the scaling dimension $h_n$ are related for general theories through 
\begin{align}\label{aii}
f_{B_3}(n)=\frac{5}{768\pi}\,\frac{h_n}{n-1}\, .
\end{align}
Now we can try to extend the calculation of $\sigma^{\ssc (6)}$ in \cite{ben2} to general R\'enyi entropies. In particular, it is natural to expect that \req{univ6} extends to general R\'enyi entropies by simply replacing $B_i$ by the $f_{B_i}(n)$ (and $A$ by some $f_A(n)$) for theories satisfying some constraint $F(f_{B_i}(n))=0$. This constraint should be such that it reduces to \req{cons6} for $n=1$. Hence, it is natural to expect a relation of the form
\begin{align}\labell{fcons6n}
3f_{B_3}(n)=f_{B_2}(n)-\frac{f_{B_1}(n)}{2}\ .
\end{align}
Let us assume that we have the appropriate constraint so that, for general values of $n$, the universal contribution to the \ren entropies becomes
\begin{align}\labell{univ7}
S_n^{\rm univ}=\log(R/\delta)\int_{\partial V} d^4y \sqrt{h}\left[2f_A(n)\, E_4 +6\pi \left(f_{B_2}(n)-\frac{f_{B_1}(n)}{4} \right)J+f_{B_3}(n)\, T_3 \right]\, .
\end{align}
Then with this structure, we find that \req{sig6} generalizes to
\begin{align}
\sigma^{\ssc (6)}_n=\frac{27\pi^3}{2}\,  f_{B_3}(n)\, .
\end{align}
Substituting \req{aii} into the above result, we find
\begin{align}\labell{track9}
\sigma^{\ssc (6)}_n=\frac{45\pi^2}{512}\,\frac{h_n}{n-1}\, ,
\end{align}
in perfect agreement with our conjecture \reef{new9}. Hence our conjecture is satisfied for the class of theories where eq.~\reef{univ7} applies.
From a different perspective, our conjecture suggests that the na\"ive generalization of \req{univ6} to \req{univ7} for general R\'enyi entropies is correct for the set of theories satisfying \req{track9}. That is, at least for these theories, the general structure of $S^{\rm univ}_{n}$ is given by \req{univ7} for general regions in six dimensions.

\section{Discussion} \labell{discuss}
In this paper, we have generalized our original conjecture \reef{conj1} \cite{bueno1,bueno2}, which relates the corner coefficient $\sigma^{\ssc (3)}$ with the central charge $\ctt$ in the two-point function of the stress tensor. Our generalization provides a similar relation for the coefficient controlling the universal contribution to the EE in almost smooth limit of a (hyper)conical singularity in the entangling surface for CFTs in general dimensions. In particular, we have shown that these cone coefficients $\sigma^{\ssc (d)}$ are again determined by $\ctt$ through
\begin{eqnarray}\label{CON1}
\sigma^{\ssc (d)}=\ctt\ \frac{\pi^{d-1}(d-1)(d-2)\Gamma[\frac{d-1}{2}]^2}{8\,\Gamma[{d}/{2}]^2\,\Gamma[d+2]}\times\left\lbrace \begin{array}{cll}
\pi&& d \text{ odd}\, ,\\ 
1 && d\, \,\text{even} \,.\end{array}\, \right.
\end{eqnarray}
We were able to prove this relation using the result of \cite{Mezei}, which applies for general holographic theories. That is, the proof applies to holographic CFTs that are dual to any higher curvature theory of gravity in the bulk, where the bulk Lagrangian does not contain derivatives of the Riemann tensor. While we have proven this relation for general holographic theories, we conjecture it applies for general CFTs in arbitrary dimensions. 

As a consistency check, we have verified that this formula is in agreement with the results obtained using the Ryu-Takayanagi  prescription \cite{rt0} for holographic theories dual to Einstein gravity in $d=3,4,6,8,10,12,14$. The odd dimensional cases with $d>3$ are more challenging. In particular, the RT prescription gives rise to very complicated implicit expressions for $a^{\ssc (d)}(\Omega)$ for $d=5, 7, \cdots$, which have proven impossible to treat analytically as yet. Interestingly, \req{CON1} provides explicit information on $a^{\ssc (d)}(\Omega)$ in these cases. It would be interesting to compute the cone coefficient $\sigma^{\ssc (d)}$ for $d=5,7,\cdots$ numerically for these holographic theories and verify that the corresponding values agree with our general formula \reef{CON1} above.

In section $\ref{renyi}$, we built on the above result to extend our conjecture \reef{conj3a} for R\'enyi corner coefficients in $d=3$ \cite{bueno3} to the following expression
 \begin{eqnarray}\labell{new10}
\sigma_n^{\ssc(d)}=\frac{h_n}{n-1}\ \frac{(d-1)(d-2)\,\pi^{\frac{d-4}{2}}\,\Gamma\left[\frac{d-1}2\right]^2}{16\ \Gamma[{d}/{2}]^3}\ \times\,\left\lbrace 
\begin{array}{cll}
\pi && d\,\, \text{odd}\,,\\
1&& d\text{ even}\,,\end{array}\, \right.
\end{eqnarray}
in general dimensions. While we have somewhat less evidence for this result, we again conjecture that it applies for general CFTs. In Table \ref{tbl2}, we summarize the theories for which this conjecture has been shown to be true so far.  These include: for $n=1$, general holographic theories in all dimensions (in the present paper and \cite{rxm}); for all $n$ for three-dimensional free scalar and fermion fields \cite{bueno1,bueno3,bueno2,Elvang,dowker}; for all $n$ for all four-dimensional theories satisfying $f_b(n)=f_c(n)$ (in the present paper and \cite{aitor9}), including \eg holographic theories and free fields; for all $n$ for all six-dimensional theories where \req{univ7} holds (in the present paper and \cite{ben2,aitor9}), including (at least for $n=1$) various holographic theories and \eg a free $\mathcal{N}=(1,0)$ hypermultiplet. We re-iterate that we have not explicitly established that \req{new10} applies for a single free scalar or fermion. It would be interesting to explore the empty slots in Table \ref{tbl2} \eg for five-dimensional and six-dimensional free scalars and fermions. Of course, a more ambitious goal would correspond to finding a complete proof of \req{new10} for general CFTs in arbitrary dimensions.


Finally, it would be interesting to further explore the implications of our results on the general structure of the 
universal contributions to R\'enyi entropy in various dimensions --- particularly in odd dimensions, for which much less is known in this respect.



\begin{table*} 
  \centering
  \begin{tabular}{c||c|c|c} 
 $d$  &   Holography & Free Fields & Constrained \\
    \hline\hline \rule{0pt}{1.2em}
$3$  & $n=1$ & $\forall\, n$&   N/A   \\ \hline\rule{0pt}{1.2em}
$4$  & $n=1$ & $\forall\, n$&   $\forall\, n$  \\ \hline\rule{0pt}{1.2em}
$5$  & $n=1$ & $-$&   N/A \\ \hline\rule{0pt}{1.2em}
$6$  & $n=1$ & $-$&  $\forall\, n$ .

  \end{tabular}
  \caption{Values of $n$ for which our generalized conjecture \req{new10} has been verified so far for $d=3,4,5,6$ theories. The first column corresponds to general holographic theories. The second makes reference to free fields (\ie a free massless scalar and fermion). The last one corresponds to $d=4$ theories for which $f_b(n)=f_c(n)$ and $d=6$ theories satisfying $3f_{B_3}(n)=f_{B_2}(n)-f_{B_1}(n)/2$ respectively. Here, `Not Applicable' for $d=3$ and 5 indicates that no analogous constrained theories have been found in these dimensions. Note that in $d=4$, at least certain holographic theories and the free fields fall in the last column but it may well apply for general CFTs. For $d=6$ and $n=1$, at least some holographic theories, but \eg not the free fields fall in the last column. }
\label{tbl2}   
\end{table*}
%



\section*{Acknowledgments}

\noindent We thank to Lorenzo Bianchi, Aitor Lewkowycz, Marco Meineri, Roger Melko, Tom\'as Ort\'in, C. S. Shahbazi,
Misha Smolkin and William Witczak-Krempa for useful discussions and comments. Research at Perimeter Institute is supported by the Government of Canada through Industry Canada and by the Province of Ontario through the Ministry of Research \& Innovation. The work of  PB has been supported by the JAE-predoc grant JAEPre 2011 00452.  
RCM acknowledges support from an NSERC Discovery grant and funding from the Canadian Institute for Advanced Research.

\appendix
 
\section{Twist operators}
\labell{Atwist}

Recall that the twist operator $\twist_n$
is defined in the replicated field theory formed as a tensor product of $n$ copies of the original QFT. In particular, it is the codimension-two surface operator extending over the entangling surface, \ie the boundary of the region $V$, whose expectation value yields 
\begin{align}
  \langle\, \twist_n\,\rangle_n = \tr\! \left[\,\rho_V^{\,n}\,\right]\,.\labell{sloppy} 
\end{align}
Here the subscript $n$ on the expectation value on the left-hand side indicates that it is taken in the $n$-fold replicated QFT. Of course, $\twist_n$ depends on the region $V$ but we have omitted  this dependence here to simplify the notation. For further discussion and details, see \eg \cite{cardy0,bueno3,holoren,twist}.
 
In the case of a CFT, the conformal scaling dimension $h_n$ of the twist operator is defined as the coefficient of the 
leading power-law divergence in the 
correlator $\langle T_{\mu\nu}\,\twist_n\rangle_n$ as the location of $T_{\mu\nu}$ approaches that of $\twist_n$ \cite{holoren,twist}. In the case of a twist operator on an infinite (hyper)plane,  this correlator is constrained by the residual conformal symmetries and conservation of the stress tensor to take the form:
\begin{align}
\langle T_{ab}\,\twist_n\rangle_n&=-\frac{h_n}{2\pi}\frac{\delta_{ab}}{y^d} \, , \qquad \langle T_{ai}\,\twist_n\rangle_n=0\,,\notag\\
\langle T_{ij}\,\twist_n\rangle_n&=\frac{h_n}{2\pi}\frac{(d-1)\delta_{ij}-d\,\hat n_i \hat n_j}{y^d} \, ,\labell{sing2} 
\end{align} 
where the indices $a,b$ and $i,j$ denote the $d$--2 parallel directions and the two transverse directions to the twist operator.\footnote{Let us add that implicitly the above expressions are normalized by dividing by $\langle \tau_n\rangle_n$ but we leave this normalization implicit 
to avoid the clutter that would otherwise be created.} Also, $y$ is the perpendicular distance from the stress tensor insertion to the twist operator and $\hat n_i$ is the unit vector orthogonally directed from $\tau_n$ to the stress tensor. 
Note that $T_{\mu\nu}$ here denotes the stress tensor for the entire $n$-fold replicated CFT. 

While the above expressions are only valid for a twist operator on a hyperplane, we stress that in general the leading singularity takes this form whenever $y\ll \ell$, 
where $\ell$ is any scale entering in the description of the geometry of the entangling surface. Hence the scaling dimension $h_n$ is a 
fixed coefficient  
which is characteristic of all twist operators $\tau_n$ (in a given CFT), independent of the details of the geometry of the corresponding entangling surface. Finally, let us add that $h_1=0$ since the twist operator $\twist_n$ becomes trivial for $n=1$.



\end{document}